\documentclass[reprint,superscriptaddress,amsmath,amssymb,aps,pra]{revtex4-2}
\usepackage{graphicx}
\usepackage{dcolumn}
\usepackage{bm}
\begin{document}
\preprint{APS/123-QED}

\title{Steady-state photoluminescence and nanoscopy of two near-identical emitters with dipole-dipole coupling}

\author{Natalia~A.~Lozing}
\affiliation{Moscow State Pedagogical University , 1-1 M Pirogovskaya St, Moscow 119991, Russia}

\author{Ekaterina~A.~Smirnova}
\affiliation{Moscow State Pedagogical University , 1-1 M Pirogovskaya St, Moscow 119991, Russia}

\author{Vladimir~K.~Roerich}
\affiliation{Microchip Technology Inc, 3870 N First St, San Jose CA,   United States 95134}

\author{Maxim~G.~Gladush}
\affiliation{Moscow State Pedagogical University , 1-1 M Pirogovskaya St, Moscow 119991, Russia}

\date{\today}

\begin{abstract}
We report progress in the theory of photoluminescence and light scattering by two closely spaced particles. This study is based on our original method to derive the master equation for a system of coupled quantum emitters driven by a cw laser. We emphasise on the case when the emitters are two-level systems but notably different in their transition frequencies and transition moments. The master equation is shown to describe the dipole-dipole coupling and the cooperative entanglement between the particles naturally. It is provided by the use of the Bogolyubov-Born-Green-Kirkwood-Yvon (BBGKY) hierarchies for reduced density matrices and correlation operators of the material and photonic subsystems. Tackling the hierarchies has also provided us with an elegant way to calculate the photoluminescence excitation spectra subject to the arrangement of the excitation and the position of the detector against the emitters. Our findings may be important for developing novel possibilities for the all-optical nanoscopy and may help construction of entangled systems to be implemented in quantum technologies.
\end{abstract}

\maketitle

\section{Introduction}
Light emission by photoluminescence from systems of interacting atom- or molecule-like particles has been the subject of extensive research in quantum optics and related areas for a few decades. It is now widely understood that ensembles of light emitters may be organized in a certain fine manner to show cooperativity and entanglement. This type of collective behaviour is a result of the interparticle interactions. The interactions produce collective excitations alternative to a sum of excited single emitters. In turn, the excitations are generally expected to damp out through some tricky scenario \cite{Dicke1954,Kuraptsev2019} and give birth to light with a few distinctive properties. The research interests are mostly associated with the super- and sub-radiance phenomena \cite{Dicke1954}. Both follow from formation of the entangled symmetric and antisymmetric eigenstates which are the fast and slow decaying collective excitations respectively. The entanglement of emitters and the dipole-dipole coupling between them are known to give rise to new resonances in the emission and absorption spectra \cite{Wodkiewicz,Agarwal1980,Richter1983,Ficek19902,Rudolph1995,Hettich2002}.
All these properties are currently of special attention in several research communities. For instance, construction of the entangled states has become a stand-alone topic in the contemporary studies for quantum computing and the quantum information science \cite{Streltsov2017,Luca2018,Lorenzo2019}. The sub-radiative states are being studied for different configurations of emitters and checked as candidates for the elements of optical and quantum memory devices \cite{Kornovan2019}.

The recent surge of interest in cooperative emission has been fueled by the advances in single molecule and single emitter spectroscopy. It allows detection and characterization of two and more fluorescing particles with subwavelength separation. Since the pioneering papers \cite{Moerner1989,Orrit1990} a number of experimental techniques have now evolved into several routines which help interdisciplinary studies. These include spectroscopy of single organic molecules \cite{Basche1992,Piliarik2014}, quantum dots \cite{Bawendi1996}, ions \cite{Sandoghdar2015}, and light-harvesting pigment-protein complexes \cite{Oijen1999}. Observation of single quantum emitters reached remarkable efficiencies and precisions shortly after it had become possible to detect zero-phonon lines (see \cite{Naumov2013} for a review). In most cases it works best at low temperatures so the vibrational excitations are ``frozen'' and the emitters can be thought of as quasi two-level systems with bright fluorescence. These spectroscopies have been so far the only successful methods to demonstrate the cooperativity of only two particles located at a small (less than the radiation wavelength) distance from one another. A demonstration of coherent optical dipole-dipole coupling of two individual molecules separated by $12$ nanometers in an organic crystal was reported in \cite{Hettich2002}. At the same time, more novel techniques to measure small separations with a fine accuracy are under way. For example, a new experimental method for direct measurements of the separation between two semiconductor colloidal quantum dots (down to $20-10$ nm) from fluorescent images was recently described in \cite{Eremchev2018}. The growing ability to observe the relationship between the distance and changes in absorption and emission requires a better theoretical description of small cooperative ensembles. Besides, recent progress in preparation of the samples makes it possible to manage and arrange particles in one-, two-, and three-dimensional structures at micro- and nano-scales. Many advanced procedures are currently practiced to successfully trap and confine small groups of emitters within a desired geometry, e.g., ions in magneto-optical traps \cite{Sauter1986,Brewer1992}, artificial atoms in a transmission line \cite{Vanloo2013}, quantum dots in liquids \cite{Ropp2010}, etc.

A system of two coupled light emitters has always been a special case of cooperativity. It has the least number of parameters which makes it a perfect object for developing a theory or interpreting an experiment. There have been numerous studies of the dynamics, entanglement, and fluorescence from paired emitters (for a review see, e.g., \cite{Ficek2002}). However, despite the long history of research, many theories had limitations. Most of the research focused on separate specific cases, which either allowed obtaining analytical expressions or simulating a particular situation. At the same time, since the early papers \cite{Milonni1975} and \cite{Ficek1987} the largest number of publications has been devoted to the study of the cooperative spontaneous decay. Because the problem is geometry-dependent, several authors have succeeded in proposing ingenious indirect ways to observe cooperativity and determine the location of emitters with respect to external conditions like in \cite{Mokhov2019}. In \cite{Redchenko2014} the time evolution of singly and doubly excited states of a system of two spatially separated qubits (two-level atoms) coupled to a one-dimensional waveguide was studied. The procedure of subwavelength optical microscopy of closely spaces particles was reported in, for example, \cite{ChangEvers2006,Evers2011,Redchenko2018} and \cite{Sun2011}. The exchange between symmetric and antisymmetric states due to the spatial variation of the applied laser excitation was studied in works \cite{Das2008,Sete2011}. In the work \citep{Akram2000} the system of two non-identical atoms showed the possibility of population trapping effect. Another direction of the research has been devoted to studying cooperative couples in presence of a continuous excitation  \cite{Lehmberg1970, AgarwalBook, Allen}. In works \cite{Richter1982,Rudolph1995} the total fluorescence intensity, fluorescence spectra and correlation functions were calculated in cases of running and standing waves, but conventionally only a case of two identical atoms was considered, and only a limited geometry of interaction was investigated. 

To date pairs of organic molecules and semiconductor quantum dots are considered being among most favored candidates for creating the smallest, best theoretically studied and most stable cooperative systems. Both types of particles can be easily imbedded into a transparent host material and fixed in unchangeable positions for unlimited time. The problem of controlling the distance between particles can be solved by preparing a variety of pairs with stochastically different separations. Controlling the distance in one pair is also a solvable problem with few theoretical restrictions -- it just requires the use of rather tricky manipulations. For such cooperative couples, two inevitable circumstances need to be considered. The first is that both types of emitters are known for their wide inhomogeneous broadening \cite{Naumov2013}. The second is that a sufficiently strong influence of the host matrix is possible \cite{Naumov2018}. Regardless of how  these circumstances manifest in a particular case the emitters in a pair cannot be considered completely identical. In this paper, we study the steady-state cooperative emission of light and how it depends on the parameters of the incident pump field. We consider a system of two near-identical two-level emitters that are driven by a cw laser beam. We derive the complete cooperative master equation for the atomic density matrix which is adjustible to the excitation geometry and appears from the conventional electric-dipole interaction Hamiltonian. The master equation is a part of the decoupled Bogolyubov-Born-Green-Kirkwood-Yvon hierarchy (BBGKY) which also provides the equations for the total photoluminescence intensity. Finally we show how a rotation of the excitation polarization may help restoring the orientation of the couple and the individual parameters of the emitters.

This paper is organized as follows. In Sec.~\ref{Sec2} we make a review of the theoretical formalism, state the problem and formulate the basic assumptions. In Sec.~\ref{Sec3} we derive the master equation in the differential-integral from and the field correlation tensor. Sec.~\ref{ME} describes reduction of the master equation to a particular solvable form. In Sec.~\ref{Sec5} we derive the equation for the total emission intensity and calculate the photoluminescence excitation spectra. We show three resonances typical for pure two-level emitter and how they change with different geometries of excitation and observation.

\section{Basic Equations and Assumptions}\label{Sec2}
\subsection{BBGKY hierarchies for emitters and photons}
The evolution of the system which includes the emitters and the field is described by a many-particle density matrix $\rho (t)$, which obeys the von Neumann equation:
\begin{equation} \label{neuman}
i\hbar \frac{d{\rho}}{dt}=[\widehat{H},{\rho}],
\end{equation}
where $\widehat{H}$ is the full Hamiltonian. We use the conventional notations for the unit imaginary number, reduced Plank constant, and commutator brackets. The Hamiltonian $\widehat{H}$ is composed of two parts $\widehat{H}=\widehat{H}_{0}+\widehat{V}$,
where $\widehat{H}_{0}$ is the energy of free particles $\widehat{H}_{0}=\sum_{i}\widehat{H}_{i}+\sum_f \widehat{H}_{f}$,  where $i$ numbers all material particles and $f$ numbers field modes,  while $\widehat{V}$ is the interaction operator.

The first step in our transformations is to replace Eq.~(\ref{neuman}) with the equivalent hierarchy of equations for reduced density matrices of the emitting system $\rho_{\lbrace a\rbrace}$, the field modes $\rho_f$, and the higher order many-particle density matrices. This can be performed using the fundamental reduction property $\rho_{\left\lbrace S\right\rbrace }=Tr_{\left\lbrace S'\right\rbrace }\rho$, where $\lbrace S\rbrace$ and $\lbrace S'\rbrace$ - arbitrary sets of elements of the system such that $\lbrace S+S'\rbrace$ is a complete set of photons and material particles. This way we obtain an infinite system of coupled equations:
\begin{equation*}
i\hbar \frac{d}{dt}\rho_{\lbrace a \rbrace} -\Big[\sum_i \widehat{H}_{i},\rho_{\small \lbrace a \rbrace} \Big]
=\sum_{f} Tr_{f}\Big[\sum_i \widehat{V}_{if},\rho_{\lbrace a \rbrace f}\Big],
\end{equation*}
\begin{equation} \label{sys1}
i\hbar \frac{d}{dt}\rho_{f} -\left[ \widehat{H}_{f},\rho_{f} \right] = Tr_{\lbrace a \rbrace}\Big[\sum_i \widehat{V}_{if},\rho_{\lbrace a \rbrace f}\Big],
\end{equation}
\begin{multline*}
i\hbar \frac{d}{dt}\rho_{\lbrace a \rbrace f} - \Big[\sum_i \widehat{H}_{i}+\widehat{H}_{f}+\sum_i \widehat{V}_{if},\rho_{\lbrace a \rbrace f} \Big] \\ =\sum_{f'\neq f} Tr_{f'}\Big[\sum_i \widehat{V}_{if'},\rho_{\lbrace a \rbrace ff'}\Big],
\end{multline*}
\begin{multline*}
i\hbar \frac{d}{dt}\rho_{ff'} - \left[ \widehat{H}_{f}+  \widehat{H}_{f'},\rho_{ff'} \right] \nonumber \\ =Tr_{\lbrace a\rbrace}\Big[\sum_i \widehat{V}_{if}+\sum_i \widehat{V}_{if'},\rho_{\lbrace a \rbrace ff'}] \\ \text{etc.}
\end{multline*}
It is advantageous to replace the many-particle density matrices with corresponding correlation operators using the cluster expansion \citep{Bonitz2016}. Each expansion is a superposition of the product of single-particle density matrices and correlation operators, satisfying $Tr_{\lbrace y \rbrace } g_{\lbrace x \rbrace}~=~0, \;\; \lbrace y \rbrace ~\in~ \lbrace x \rbrace$, as follows:
\begin{eqnarray*}
\rho_{ff'} = \rho_f \rho_{f'} + g_{ff'}, \;\;\;
\rho_{\lbrace a \rbrace f}=\rho_{\lbrace a \rbrace } \rho_{f} +g_{\lbrace a \rbrace f}, \\
\rho_{\lbrace a \rbrace ff'}=\rho_{\lbrace a \rbrace } \rho_{f} \rho_{f'}+\rho_{\lbrace a \rbrace }g_{ff'}
+\rho_{f}g_{\lbrace a \rbrace f'}\\
+\rho_{f'}g_{\lbrace a \rbrace f}+g_{\lbrace a \rbrace ff'}\\ \text{etc.}
\end{eqnarray*}
Substituting the expansions into Eq.~(\ref{sys1}) produces the hierarchy with one emitter equation
\begin{equation} \label{atom1}
i\hbar \frac{d}{dt}\rho_{\lbrace a \rbrace } -\Big[\sum_i \overline{{H}}_{i},\rho_{\lbrace a \rbrace } \Big] =\sum_{f} Tr_{f}\Big[\sum_i \widehat{V}_{if},g_{\lbrace a \rbrace f}\Big]
\end{equation}
and an infinite number of equations for the field modes
\begin{equation} \label{field1}
i\hbar \frac{d}{dt}\rho_{f} -\left[ \overline{H}_{f},\rho_{f} \right] = Tr_{\lbrace a \rbrace }\Big[\sum_i \widehat{V}_{if},g_{\lbrace a \rbrace f}\Big].
\end{equation}
The left hand sides of Eqs.~(\ref{atom1}) and (\ref{field1}) contain the effective energy operators. They are the sums of the energy of free particles and Hartree potentials, which are partial quantum-mechanical averages $\langle \widehat{O} \rangle_S =Tr_S \widehat{O} \rho_S $ with respect to particular species:
\begin{eqnarray*}
\overline{H}_{i}=\widehat{H}_{i}+\sum_{f} \langle \widehat{V}_{if} \rangle_{f}, \\
\overline{H}_{f}=\widehat{H}_{f}+ \sum_i \langle \widehat{V}_{if} \rangle_{i}.
\end{eqnarray*}
These Hartree potentials provide the first-order coupling between the material particles and the field. In Eq.~(\ref{atom1}) one finds the energy of interaction between each emitter and the mean electric field while in equation (\ref{field1}) this potential refers to interaction of each field mode with the mean induced dipole. The second order coupling arises from the right hand sides where both equations contain the emitters-field correlation $g_{\lbrace a \rbrace f}$. The equation for $g_{\lbrace a \rbrace f}$ follows from the similar algebra applied to Eq.~(\ref{sys1}) and the cluster expansion. We write it in a form convenient for further transformations:
\begin{multline} \label{corA}
i\hbar \frac{d}{dt}g_{\lbrace a \rbrace f}- \left[ \sum_i \overline{H}_{i} + \overline{H}_{f}, g_{\lbrace a \rbrace f} \right]+\widehat{S}
= \widehat{I}+ \widehat{L}+\widehat{\Pi}+\widehat{C}.
\end{multline}
The commutator in the left hand side contains the same energy operators as in Eqs.~(\ref{atom1}) and (\ref{field1}).  Operator $\widehat{S}$ combines corrections to the Hartree terms which are three commutators:
\begin{eqnarray*} \begin{split}
\widehat{S}=&\left[ \sum_i \langle \widehat{V}_{if} \rangle_{i}+\langle \widehat{V}_{if} \rangle_{f}, g_{\lbrace a \rbrace f} \right]+\rho_{f}Tr_{f} \left[\sum_i \widehat{V}_{if} ,g_{\lbrace a \rbrace f} \right] \\ \nonumber
&+\rho_{\lbrace a \rbrace }Tr_{\lbrace a \rbrace } \left[ \sum_i \widehat{V}_{if},g_{\lbrace a \rbrace f} \right].
\end{split}\end{eqnarray*}
The four right hand side operators represent the hierarchical structure of how the emitter-field correlation is coupled to the density matrices and the higher order correlations:
\begin{eqnarray*} \begin{split}
\widehat{I}=&\left[ \sum_i \overline{V}_{if}, \rho_{\lbrace a \rbrace }\rho_{f} \right], \nonumber \\
\widehat{L}=&\left[ \sum_i {V}_{if}, g_{\lbrace a \rbrace f} \right] ,\\ \nonumber
 \widehat{\Pi} =&\sum_{f' \neq f} Tr_{f'} \left[ \sum_i \widehat{V}_{if},\rho_{\lbrace a \rbrace }g_{ff'} \right], \\
\widehat{C}=&\sum_{f' \neq f} Tr_{f'} \left[\sum_i \widehat{V}_{if},g_{\lbrace a \rbrace ff'} \right].
\end{split}\end{eqnarray*}
This hierarchy of couplings corresponds to the formal representation of the BBGKY described in \citep{Bonitz2016}.

The first operator $ \widehat{I}$ is a pure inhomogeneity formed by the operators of the orders lower than $g_{\lbrace a \rbrace f}$, i.e., the product of the reduced density matrices. It contains the interaction operators $V_{if}$ corrected with the corresponding components of the Hartree potentials:
\begin{equation*}
{\overline{V}}_{if} = \widehat{V}_{if}-\langle \widehat{V}_{if} \rangle_{i}-\langle \widehat{V}_{if} \rangle_{f}.
\end{equation*}
Operator $\widehat{L}$ is homogeneous with respect to $g_{\lbrace a \rbrace f}$ and is known to collect so-called ladder terms. These are the commutators between the quantum mechanical interactions of the emitters with the field mode of the attention and the correlation operator. Obviously this term is of the higher order as compared to the left hand side of Eq.~(\ref{corA}). The other operators are inhomogeneous terms which provide the links to the other types of correlations. Operator $\widehat{\Pi}$ is a collection of the terms that describe contributions from the correlations between two different field modes. The term $\widehat{C}$ is the highest order inhomogeneity, i.e., contains the three particle correlation which is the bridge to the rest of the BBGKY.

Equation~(\ref{corA}) suggests the photon-photon correlations must be included. The corresponding equations follow the above structure and are the following:
\begin{equation} \label{corF}
i\hbar \frac{d}{dt}g_{ff'} -\left[ \overline{H}_{f}+\overline{H}_{f'},g_{ff'}
 \right] = \widehat{\Pi}' + \widehat{C}' ,
\end{equation}
where
\begin{eqnarray} \nonumber
\widehat{\Pi}'= Tr_{\lbrace a \rbrace } &&\left[\sum_i \widehat{V}_{if},\rho_f g_{\lbrace a \rbrace f'} \right]\\ \nonumber
&&+ Tr_{\lbrace a \rbrace } \left[\sum_i \widehat{V}_{if'},\rho_{f'}g_{\lbrace a \rbrace f} \right],
\end{eqnarray}
\begin{equation*}
\widehat{C}'= Tr_{\lbrace a \rbrace }\bigg[\sum_i \widehat{V}_{if} + \sum_i \widehat{V}_{if'},g_{\lbrace a \rbrace ff'}\bigg].
\end{equation*}
For the purposes of this study it is sufficient to remain within the second order operators in the inhomogeneous terms and in the low order homogeneous terms, that is, we set $\widehat{L}=\widehat{C} = \widehat{C}'=0$ and disregard the impact of contributions of all higher order correlations between the subsystems.

It means we go further beyond the Born approximation and take the advantage of building the hierarchy either in polarization or ladder approximation. In either case we have a closed system of equations for reduced density and correlation matrices, and, consequently, can proceed with solving a self-consistent problem by stating the necessary initial conditions.

Further it is beneficial to transform to the interaction picture in which the total density matrix is
\begin{equation*}
\widetilde{\rho}=exp \left\lbrace \frac{i}{\hbar} \widehat{H}_{0} t \right\rbrace \rho \: exp \left\lbrace -\frac{i}{\hbar} \widehat{H}_{0}t \right\rbrace.
\end{equation*}
Hereinafter, we will write the operators without the tildes. Now we have come to a general closed system of equations for density matrices and correlation operators which we will use in our analysis:
\begin{multline} \label{roA}
i\hbar \frac{d}{dt} \rho_{\lbrace a \rbrace } - \sum_{f}\left[\sum_i  \langle \widehat{V}_{if}\rangle_{f},\rho_{\lbrace a \rbrace } \right] \\
 = \sum_{f} Tr_{f} \left[ \sum_i \widehat{V}_{if}, g_{\lbrace a \rbrace f} \right],
\end{multline}
\begin{equation} \label{roF}
i\hbar \frac{d}{dt} \rho_{f} -  \left[ \sum_i \langle \widehat{V}_{if}\rangle_{i} ,\rho_{f} \right] =  Tr_{\lbrace a \rbrace }\left[\sum_i\widehat{V}_{if}, g_{\lbrace a \rbrace f} \right],
\end{equation}
\begin{multline} \label{gAAF}
i\hbar \frac{d}{dt}g_{\lbrace a \rbrace f}
= \left[\sum_i \overline{V}_{if}, \rho_{\lbrace a \rbrace }\rho_{f} \right] \\+ \sum_{f' \neq f} Tr_{f'} \left[\sum_i \widehat{V}_{if'}, \rho_{\lbrace a \rbrace } g_{ff'} \right],
\end{multline}
\begin{multline} \label{gFF}
i\hbar \frac{d}{dt} g_{ff'} - \left[ \sum_i  \langle \widehat{V}_{if}  +  \widehat{V}_{if'} \rangle_i ,g_{ff'} \right]  \\
= Tr_{\lbrace a \rbrace }  \left[ \sum_i\widehat{V}_{if},\rho_f g_{\lbrace a \rbrace f'} \right]
+Tr_{\lbrace a \rbrace }  \left[\sum_i \widehat{V}_{if'},\rho_{f'}g_{\lbrace a \rbrace f} \right].
\end{multline}
In this system Eq.~(\ref{roA}) is unique while Eqs.~(\ref{roF})-(\ref{gFF}) are multiple as they refer to the photonic modes. Our further developments will be introducing formal solutions of Eqs.~(\ref{roF})-(\ref{gFF}) into (\ref{roA}).

\subsection{Statement of the problem for two-particle systems}\label{2b}
\begin{figure}
  \includegraphics[width=0.48\textwidth]{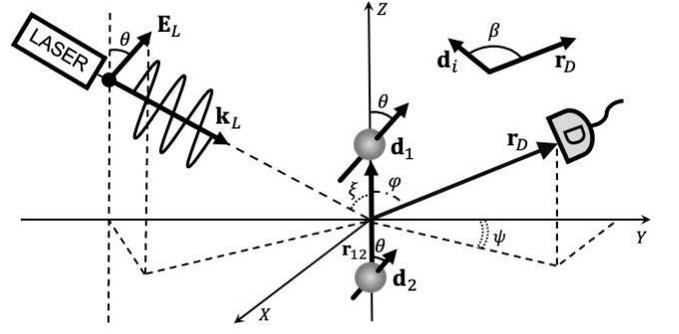}
  \caption{The excitation/detection diagram. Two emitters are fastened at a distance $r_{12}=\vert{\bf r}_{12}\vert$ apart. The excitation is described by vectors ${\bf k}_L$ and ${\bf E}_L$ (${\bf E}_L\perp {\bf k}_L$) and is adjustable by varying polar angles $\xi$ and $\theta$. The dipole moments ${\bf d}_i$ are defined by polar angle $\theta$. Detector position is defined by ${\bf r}_D$ with polar $\varphi$ and azimuthal $\psi$ angled. The angle between ${\bf r}_D$ and ${\bf d}_i$ is $\beta$.}\label{Fig1}
\end{figure}

In this work we consider a pair of quantum emitters excited by a cw laser light. Each emitter is approximated as a two-level atom with a transition frequency $\omega_i$ and a transition dipole moment ${\bf d}_i$, where $i=1,2$, such that $\omega_1 \ne \omega_{2}$, $\vert{\bf d}_1\vert \ne \vert{\bf d}_{2}\vert$ and $\vert\omega_1 - \omega_{2}\vert \ll \omega_1, \omega_{2}$. The excitation light beam may be, in principle, of arbitrary composition but in this paper as a convention assumption for a cw laser it is treated as a linearly polarized plane wave with the wave vector ${\bf k}_L$ and the electric field strength amplitude ${\bf E}_L$. The atoms are viewed as attached to static positions separated by a distance represented by a radius-vector ${\bf r}_{12}$, that is a parameter to change arbitrary. The photoluminescence of the atomic pair is to be calculated for arbitrary directions ${\bf r}_{D}$ from the midpoint of the system.
Figure~\ref{Fig1} illustrates this statement of the problem and possible arrangements for the excitation and detection means. In this diagram we set the Cartesian system such the emitters are placed along the $Z$-axis at equal distances from the origin. The spatial arrangement (relative orientation) for the excitation system ${\bf k}_L$ and ${\bf E}_L$, vectors ${\bf r}_{12}$ and ${\bf r}_D$ may be fully defined by setting independent values  to four angles: $\xi$, $\theta$, $\varphi$, and $\psi$. The excitation scheme and the directions of induced dipole moments are determined by two polar angles $\xi$ and $\theta$ that refer to ${\bf k}_L$ and ${\bf E}_L$ orientations respectively. The azimuthal position of ${\bf d}_i$ (not marked on the figure) is a function $f(\xi,\theta)$ which is calculated given that ${\bf d}_i$ and ${\bf E}_L$ are collinear and ${\bf k}_L$ lies in the $ZY$-plane. The inclination of ${\bf d}_i$ is dependent on $\xi$ and ranges within $\vert\pi/2-\xi\vert \leq \theta \leq \pi-\vert\pi/2-\xi\vert$. Direction to the observer ${\bf r}_D$ may be varied by  hand defining the values for polar and azimuthal angles $\varphi$ and $\psi$. Consequently, we have sufficient data to calculate the angle $\beta(\xi,\theta,\varphi,\psi)$ between vectors ${\bf r}_D$ and ${\bf d}_i$ that we refer to as the `observation parameter'.

The free Hamiltonian of the system is:
\begin{multline*}
\widehat{H}_{0}=\sum_{i}\widehat{H}_{i}+\sum_f \widehat{H}_{f}=\sum_i \hbar\omega_{i}\hat{\sigma}^+_{i}\hat{\sigma}^-_{i} + \sum_{f} \hbar \omega_{f} \hat{a}^{\dagger}_{f} \hat{a}_{f},
\end{multline*}
where we have two types of particle species, i.e., the emitters and the field photons. The first sum represents the energies of free material particles with raising $\hat{\sigma}_i^+=\vert e_i \rangle \langle g_i \vert$ and lowering $\hat{\sigma}_i^-=\vert g_i \rangle \langle e_i \vert$ operators, where $\vert e_i \rangle$ and $\vert g_i \rangle$ are the eigenvectors for the excited and the ground states, index '$i$' numbers all the emitters ($i\in\lbrace{1, 2}\rbrace$).  The second sum represents the energy of a free quantized electromagnetic field, $\hat{a}_f^{\dagger}$ and $\hat{a}_f^-$ are photon creation and annihilation operators respectively. Index $f=\lbrace {\bf k},\alpha \rbrace$ determines the mode frequency and direction (${\bf k}_f$ - wave vector, $ | {\bf k}_f | = \omega_f /c$) and polarization $(\alpha \in\lbrace 1,2 \rbrace)$.

Operator $\widehat{V}$ describes the interaction of each emitter with electromagnetic field modes in the electric dipole approximation:
\begin{equation*}
\widehat{V}=\sum_{f}\sum_i \widehat{V}_{if}, \;\;\;\;\ \widehat{V}_{if}= - \widehat{\bf d}_{i} \cdot \widehat{\bf E}_{f},
\end{equation*}
where $\widehat{\bf d}_i$ – quantum dipole moment operator:
\begin{equation*}
\widehat{\bf d}_{i} = \hat{\sigma}^-_{i} {\bf d}_{i} + \hat{\sigma}^+_{i} {\bf d}^{*}_{i},
\end{equation*}
and $ \widehat{\bf E}_f$ - quantized electric field mode operator:
\begin{equation*}
\widehat{\bf E}_{f}\left( {\bf r} \right)=i \lambda_{f} {\bf e}_{f} \left(\hat{a}_{f} e^{i{\bf k}_f\cdot {\bf r}}-\hat{a}_{f}^\dagger e^{-i{\bf k}_f \cdot {\bf r}}\right),
\end{equation*}
where ${\bf e}_f$ – unit polarization vector, $\lambda_f=\sqrt{ 2 \pi \hbar \omega_f/V}$ –  coupling constant in the quantization volume $V$.

In the interaction representation the  dipole moment operator and the electric field mode operator acquire an explicit time dependencies:
\begin{equation*}
\widehat{\bf d}_{i} (t) = \hat{\sigma}^-_{i} {\bf d}_{i} e^{-i\omega_{i}t}+ \hat{\sigma}^+_{i} {\bf d}^{*}_{i}e^{i\omega_{i}t},
\end{equation*}
\begin{equation*}
\widehat{\bf E}_{f} ({\bf r},t) = i\lambda_{f} {\bf e}_{f} \left(\hat{a}_{f}e^{-i\omega_{f}t+i{\bf k}_f\cdot{\bf r}}-\hat{a}^{\dagger}_{f} e^{i\omega_{f}t-i{\bf k}_f\cdot{\bf r}}\right).
\end{equation*}

\section{Equations for emitters and the field-tensor}\label{Sec3}
Equation~(\ref{roA}) is the one of our attention and will be transformed into a master equation. To do this, we first rearrange the terms and write it in the form:
\begin{multline} \label{ro}
\frac{d}{dt} \rho_{\lbrace a \rbrace }=\frac{i}{\hbar} \sum_i \left[
		\widehat{\bf d}_{i}(t)\cdot \bm{ \mathcal{E}} ({\bf r}_{i},t), \rho_{\lbrace a \rbrace }(t)\right] \\
		 +\frac{i}{\hbar} \sum_i \left[\widehat{\bf d}_{i}(t)\cdot \widehat{\bf\Phi}({\bf r}_{i},t)\right],
\end{multline}
where the commutator of the vector operators reads as $[\hat{\bf a}\cdot\hat{\bf b}]=(\hat{\bf a}\cdot\hat{\bf b})-(\hat{\bf b}\cdot\hat{\bf a})$. The traces in Eq.~(\ref{roA}) have produced the quantum mechanical averages with respect to the field operators which we denote as $\bm{ \mathcal{E}}({\bf r},t)$ and $\widehat{\bf \Phi}({\bf r},t)$:
\begin{equation} \label{E}
\bm{ \mathcal{E}}({\bf r},t)=\sum_{f}Tr_f \lbrace \widehat{\bf E}_f ({\bf r},t) \rho_f(t) \rbrace,
\end{equation}
\begin{equation} \label{X}
\widehat{\bf \Phi}({\bf r},t)=\sum_{f}Tr_f \lbrace \widehat{\bf E}_f ({\bf r},t) g_{\lbrace a \rbrace f}(t) \rbrace.
\end{equation}
Once evaluated at the positions of the emitters ${\bf r}_i$ they represent respectively the action of the effective mean field and weaker contributions from the emitter-field correlations. Equations~(\ref{E}) and (\ref{X}) should be expanded through substitution of the formal solutions of Eqs.~(\ref{roF})-(\ref{gFF}). This will produce the commutators of the the field operators:
\begin{equation} \label{Grin}
\sum_{f} \left[ \widehat{\bf E}_f({\bf r},t)\otimes\widehat{\bf E}_f({\bf r}',t')
 \right]=-i\hbar \overleftrightarrow{\bf G} ({\bf r} - {\bf r}', t-t'),
\end{equation}
where $ \overleftrightarrow {\bf G} ({\bf r},t) $ is the Green's tensor of electromagnetic field propagation in vacuum \citep{Mandel}. Note that the commutation in this case must be treated as $\big[\hat{\bf a}\otimes\hat{\bf b}\big]=\hat{\bf a}\otimes\hat{\bf b}-(\hat{\bf b}\otimes\hat{\bf a})^T$ and give a dyadic product.

For the positive and negative frequency components of the electric field $\widehat{\bf E}_f = \widehat{\bf E}_f^{+} + \widehat{\bf E}_f^{-} $ we use the following commutation relations:
\begin{equation} \label{GrinRet}
\sum_f \left[ \widehat{\bf E}_f^{\pm} ({\bf r},t)\otimes \widehat{\bf E}^{\mp}_f({\bf r}_i,t') \right] = -i\hbar \overleftrightarrow{\bf G}^{\pm} ({\bf r}-{\bf r}_i,t-t'),
\end{equation}
where $\overleftrightarrow{\bf G}^{\pm}$ - advanced and retarded parts of the Green's tensor.

Consequently, for equation (\ref{E}) we find:
\begin{equation} \label{Efull}
\bm{\mathcal{E}}({\bf r},t)=\bm{ \mathcal{E}}_0({\bf r},t)+ \int_0^{t} dt' \sum_i \overleftrightarrow{\bf G} ({\bf r}-{\bf r}_i,t-t'){\bf p}_i(t'),
\end{equation}
where $ \bm{\mathcal{E}}_0 ({\bf r},t) = \sum_{f}Tr_f \lbrace \widehat{\bf E}_f ({\bf r},t) \rho_f^0 \rbrace$ is determined by the initial value of $\rho_f^0=\rho_f(t=0)$.  This part of the field describes all possible incident perturbations acting on the emitters as well as noise background.  We may assume a situation when at $t=0$ the quantized field is in the vacuum state except for one or a few selected modes in coherent states.  This assumption will set up the presence of the incident laser beam.  Terms ${\bf p}_i(t)=\langle \widehat{\bf d}_i(t) \rangle_i$ are namely the induced dipole moments, so the second part in (\ref{Efull}) represents the reaction field.

Using a formal solution of the equation (\ref{gAAF}) and commutation relations of field operators (\ref{GrinRet}), we obtain the expression for $\widehat{\bf \Phi}$:
\begin{multline} \label{Phi}
\widehat{\bf \Phi} ({\bf r},t) = \sum_i\int_0^{t} dt' \bigg \lbrace \overleftrightarrow{\bf G}^{+} ({\bf r}-{\bf r}_i,t-t')\overline{\bf d}_i (t') \rho_{\lbrace a \rbrace }(t') \\
+ \overleftrightarrow{\bf G}^{-} ({\bf r}-{\bf r}_i,t-t') \rho_{\lbrace a \rbrace }(t') \overline{\bf d}_i (t') \bigg \rbrace \\
+ \frac{i}{\hbar} \sum_i  \int_0^t dt'\overleftrightarrow {\bf W}({\bf r},{\bf r}_i,t,t')\left[ \widehat{\bf d}_i(t'),\rho_{\lbrace a \rbrace }(t') \right].
\end{multline}
The first sum in this equation contains retarded Green's tensors and $\overline {\bf d}_i = \widehat{\bf d}_i - {\bf p}_i$. The terms in this sum are evaluated with previous states of the emitters, i.e.,  $\rho_{\lbrace a \rbrace}(t'),  t'\leq t$. The second sum contains tensor we have denoted as $\overleftrightarrow {\bf W}$ and further refer to as the scattered field correlation tensor:
\begin{multline} \label{W1}
\overleftrightarrow {\bf W}({\bf r},{\bf r}',t,t')=\sum_{f, f' \neq f} \Big\lbrace \langle : \widehat{\bf E}_f ({\bf r},t) \otimes \widehat{\bf E}_f ({\bf r}',t'): \rangle_f  \\
 -\langle \widehat{\bf E}_f ({\bf r},t) \rangle_f \otimes \langle \widehat{\bf E}_f ({\bf r}',t') \rangle_f  \\
 +  Tr_{ff'} \lbrace\widehat{\bf E}_f ({\bf r},t) \otimes \widehat{\bf E}_{f'} ({\bf r}',t') g_{ff'} (t') \rbrace\Big\rbrace,
\end{multline}
where : : means the normal ordering of the operators. This tensor describes the correlation of the components of nonclassical incoherent electromagnetic field and the dynamic intermode redistribution of photons.  Note that for a field described by a set of coherent states, this tensor will be strictly equal zero. It must be outlined that this tensor is dependent on the previous performance of the scatterers (the two emitters in our case).  Indeed, in (\ref{W1}) one can see that $\rho_f$ and $g_{ff'}$ are linked with $\rho_{\lbrace a \rbrace}$ as it follows from the equations (\ref{roF})-(\ref{gFF}). It means these terms in the emitter-field correlation provides the feedback loop via the scattered field.

If the time required for the light signal to transverse the system is small in the comparison to the time $\Delta t$ required for appreciable changes in population of the atomic levels, it is possible to use Markov approximation, i.e., to put $\rho(t')~\approx~\rho(t)$. For the interatomic distances of the order of the resonant wavelength this condition is satisfied.

Now we need to rewrite the material equation using the above results.
Thus, introducing the formal solution of (\ref{roF}) and (\ref{gAAF}) into (\ref{roA}) and using Markov approximation equation (\ref{roA}) transforms into:
\begin{multline} \label{roa}
\frac{d}{dt} \rho_{\lbrace a \rbrace } = 	\frac{i}{\hbar} \sum_i \left[
		\widehat{\bf d}_{i}(t) \bm{ \mathcal{E}_0} ({\bf r}_{i},t), \rho_{\lbrace a \rbrace }\right] \\
		+\frac{i}{\hbar} \sum_{i,j} \Bigg[\widehat{\bf d}_{i}(t),\int_0^{t} dt' \bigg \lbrace \overleftrightarrow{\bf G}^{+} ({\bf r}_i-{\bf r}_j,t-t'){\bf d}_j (t') \rho_{\lbrace a \rbrace }\\
			+ \overleftrightarrow{\bf G}^{-} ({\bf r}_i-{\bf r}_j,t-t') \rho_{\lbrace a \rbrace } {\bf d}_j (t')\bigg\rbrace \Bigg]\\
		-\frac{1}{\hbar^2} \sum_{i,j} \left[\widehat{\bf d}_{i}(t), \int_0^t dt'\overleftrightarrow {\bf W}({\bf r}_i,{\bf r}_j,t,t')\left[ \widehat{\bf d}_j(t'),\rho_{\lbrace a \rbrace } \right]\right].
\end{multline}
The first group of commutators in the RHS contains the field $ \bm{\mathcal{E}}_0$ from equation (\ref{Efull}),  which in a particular case may represent the incident laser.  One can immediately recognize the appearance of the relevant Rabi frequencies, that fulfil the coherent part of the interaction.  Note that the time integral in (\ref{Efull}) cancels with the terms with ${\bf p}_i$ from (\ref{Phi}) under Markov approximation. That means a natural exclusion of the self-action of the emitters provided by BBGKY.

The second sum collects the commutators which are responsible for various incoherent interactions (energy exchange) between the emitters and the field.  Because of the properties of the Green's tensor these terms can be separated with respect to its real and imaginary parts. Besides, each part will have the commutations in which the Green's tensor is taken at the poles (when $i=j$) and the commutations in which it is evaluated for the distances between the emitters ($i\ne j$). Thus, one may anticipate these term will describe the radiative decay and transition frequency shifts respectfully.

The last group of terms contains the field correlation tensor (\ref{W1}). The operator part suggests that they will produce multiple cross-contributions to the kinetics of the populations and coherencies in the emitting pair. These contributions are presumably characterized by the rates determined by the flux density of incoherent photons. That includes multiple scattering of light and excitation exchange between the emitters.

\section{The cooperative master equation} \label{ME}
To get the solvable equation for $\rho_{\lbrace a \rbrace }$ we use the rotating wave approximation (RWA) and go back to Schroedinger's picture into the frame rotating with the laser frequency.  As driving field we consider the plane monochromatic wave $ \bm{ \mathcal E}_0 ({\bf r}, t) = Re( \bm{\mathcal E} e^{i\omega_L t - i {\bf k}_L \cdot {\bf r}}) $ with complex amplitude $ \bm{ \mathcal{E}} $,  frequency $ \omega_L $ and wave vector $ {\bf k}_L $.
Thus,  we perform the following unitary transformation:
\begin{equation*}
\widetilde{\widetilde{\rho}}=exp \left\lbrace i \sum_{i} \Delta_i \hat{\sigma}^+_i \hat{\sigma}^-_i t \right\rbrace \rho \: exp \left\lbrace -i \sum_{i} \Delta_i\hat{\sigma}^+_i \hat{\sigma}^-_i t \right\rbrace.
\end{equation*}
where $\Delta_i=\omega_i-\omega_L$ is the detuning of the laser frequency from the corresponding transition frequencies of the emitters.

The time integral in the second sum in (\ref{roa}) are computable as the a one-way Fourier transform of the Green's function:
\begin{eqnarray*}
\int_0^t dt' \overleftrightarrow{\bf G}^\pm ({\bf r},t-t')e^{i \omega (t-t')} = \overleftrightarrow{\bf G}^\pm ({\bf r},\omega).
\end{eqnarray*}
It satisfy the following relations:
\begin{eqnarray*}
&&\overleftrightarrow{\bf G}^- ({\bf r},\omega)=0, \;\;\; \overleftrightarrow{\bf G}^+ ({\bf r},-\omega)=0,\\
&&\overleftrightarrow{\bf G}^+ ({\bf r},\omega)= \overleftrightarrow{\bf G} ({\bf r},\omega)\\
&&\overleftrightarrow{\bf G}^- ({\bf r},-\omega)=\overleftrightarrow{\bf G} ({\bf r},-\omega)=\overleftrightarrow{\bf G}^{*} ({\bf r},\omega).
\end{eqnarray*}
\begin{figure}
  \includegraphics[width=0.4\textwidth]{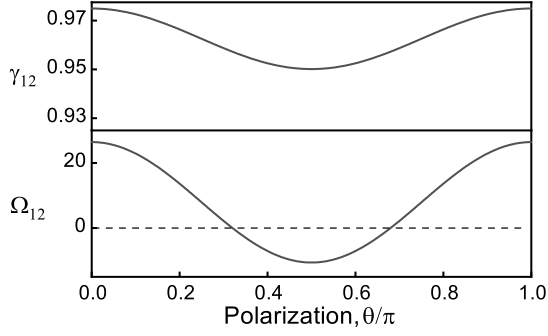}
  \caption{The cooperative relaxation rate (upper) and the parameter of dipole-dipole interaction (lower) as functions of angle $\theta$ for constant separation $r_{12}=0.08\lambda$ ($\lambda=\lambda_1$). }\label{Fig8}
\end{figure}

In considered picture after integration this sum consists of the terms we denote as $Z_{ij}$, which take the following form:
\begin{equation*}
Z_{ij} = D_{ij} \left[ \hat{\sigma}_i^{+},\hat{\sigma}_j^{-} \rho_{\lbrace a \rbrace } \right] + D_{ij}^{*} \left[ \hat{\sigma}_i^{-},\rho_{\lbrace a \rbrace } \hat{\sigma}_j^{+} \right],
\end{equation*}
where
\begin{equation*}
D_{ij}=\frac{1}{\hbar} {\bf d}_i^{*} \overleftrightarrow{\bf G} ({\bf r}_i - {\bf r}_j, \omega_0) {\bf d}_j.
\end{equation*}
Here, $\omega_0=(\omega_1+ \omega_2)/2$ is the emitters average frequency.

The imaginary part of $ Z_{ij} $ is responsible for the relaxation of the density matrix:
\begin{equation*}
\mathrm{Im} Z_{ij} = \gamma_{ij} \left( \left[\hat{\sigma}_i^{+},\hat{\sigma}_j^{-}\rho_{\lbrace a \rbrace } \right]+\left[\rho_{\lbrace a \rbrace }\hat{\sigma}_i^{+},\hat{\sigma}_j^{-} \right] \right),
\end{equation*}
where $\gamma_{ij}=\mathrm{Im}(D_{ij} )$. For $ {\bf d}_{i}^* = {\bf d}_{i}$, we get:
\begin{equation*}
\gamma_{ij} =\frac{1}{\hbar} {\bf d}_i \mathrm{Im} \left( \overleftrightarrow{\bf G} ({\bf r}_i - {\bf r}_j, \omega_0) \right) {\bf d}_j.
\end{equation*}
If $ i = j $, then $\gamma_{ii}=\gamma_i=2|{\bf d}_i |^2 \omega_i^3/(3 \hbar c^3)$ is the radiation relaxation rate of a single emitters; for $ i \neq j $, the value $ \gamma_{ij} $ determines the collective damping rate.

The real part of $Z_{ij}$ produces the radiative shifts for $i=j$ which refer to the transitions of individual emitters while for $i\ne j$ we get the shifts of the energy levels of the collective ensemble:
\begin{equation*}
\mathrm{Re} Z_{ij} =  \Omega_{ij} \left[ \hat{\sigma}_i^{+} \hat{\sigma}_j^{-}, \rho_{\lbrace a \rbrace } \right] ,
\end{equation*}
where
\begin{equation*}
\Omega_{ij} = \mathrm{Re}(D_{ij}) = \frac{1}{\hbar} {\bf d}_i \mathrm{Re} \left( \overleftrightarrow{\bf G} ({\bf r}_i - {\bf r}_j, \omega_0) \right) {\bf d}_j
\end{equation*}
is the individual radiative shift for $i=j$ and the dipole-dipole interaction parameter for $i\ne j$.
In the current work we will neglect the individual radiative shifts. Taking these terms into account leads to small changes in the transition frequencies ($\omega_i \rightarrow \omega_i + \Omega_{ii}$).  Obviously, we may rely on the convention to assume these shifts as included to $\Delta_i$.

To calculate the collective relaxation rate and the dipole – dipole interaction parameter,  we use the explicit form of the Green's tensor of the electromagnetic field propagation:
\begin{eqnarray*}
\overleftrightarrow{\bf G} ({\bf r}_{ij},\omega) =k^2 \frac{e^{ikr_{ij}}}{r_{ij}} \left( R(ikr_{ij})\overleftrightarrow{\bf I}+Q(ikr_{ij}) {\bf n}_{ij} \otimes {\bf n}_{ij}\right),\\
{\bf r}_{ij} = {\bf r}_i - {\bf r}_j, \; \;\;{r}_{ij}=\vert {\bf r}_{ij} \vert, \;  \;\; k=\frac{\omega}{c}, \; \; \; {\bf n}_{ij} \otimes {\bf n}_{ij}=\frac{{\bf r}_{ij} \otimes {\bf r}_{ij}}{{\bf r}_{ij}^2},\\
P(z) = 1 - \frac{1}{z} + \frac{1}{z^2}, \; \; \;\; Q(z)=-1+\frac{3}{z}-\frac{3}{z^2}.\; \; \;\;\; \; \;\;
\end{eqnarray*}

Thus, the relaxation rate and dipole-dipole interaction parameters take the known form \cite{Milonni1975, Akram2000}:
\begin{multline*}
\gamma_{12}=\gamma_{21}=\frac{3}{2} \frac{\sqrt{\gamma_{1}\gamma_2}}{k_0 r_{12}} \Bigg( \cos (k_0 r_{12})\left[ \frac{1}{k_0 r_{12} }- \frac{3\cos^2 \theta}{k_0 r_{12}} \right] \\
 + \sin ( k_0 r_{12}) \left[ 1 -\cos^2\theta + \frac{3\cos^2 \theta - 1}{k_0^2r_{12}^2} \right] \Bigg),
\end{multline*}
and
\begin{multline*}
\Omega_{12}=\Omega_{21}=\frac{3}{2} \frac{\sqrt{\gamma_{1}\gamma_2}}{k_0 r_{12}} \Bigg( \sin(k_0 r_{12}) \left[  \frac{3\cos^2 \theta}{k_0 r_{12}} -\frac{1}{k_0 r_{12} }\right]\\ +\cos(k_0 r_{12})\left[ 1 -\cos^2\theta + \frac{3\cos^2 \theta - 1}{k_0^2r_{12}^2} \right]
  \Bigg),
\end{multline*}
Here, $ \theta $ is the angle between the direction of the dipole moments $ {\bf d}_i $ and the vector connecting two emitters $ {\bf r}_{12} $, and $k_0=(k_1+k_2)/2$.
The relaxation rate and dipole-dipole interaction parameters describe the collective damping and the collective shift of energy levels and determine the collective properties of the two-atom system.  The interaction strongly depends on interatomic separation $r_{12}$ and the spatial orientation of dipole moments.  Figure \ref{Fig8} illustrates the dependence of the collective parameters $\gamma_{12}$ and $\Omega_{12}$ (hereinafter all parameters are scaled to $\gamma_{1}$ and,  for simplicity, we take $\gamma_{1}=1$) on the angle $\theta$, which reflects the incident field polarization (see Fig.\ref{Fig1}).

Finally, we obtain the master equation describing the time evolution of the atomic system:
\begin{widetext}
\begin{multline} \label{roAA}
\frac{d}{dt} \rho_{\lbrace a \rbrace } =
-i \sum_i \Delta_i \left[ \hat{\sigma}^+_i \hat{\sigma}^-_i, \rho_{\lbrace a \rbrace } \right] + i \sum_i \left[\Omega_i({\bf r}_i)\hat{\sigma}^-_i+\Omega^*_i({\bf r}_i) \hat{\sigma}^+_i,\rho_{\lbrace a \rbrace } \right]
		+ i\sum_{i\ne j} \Omega_{ij} \left[ \hat{\sigma}_i^{+} \hat{\sigma}_j^{-}, \rho_{\lbrace a \rbrace } \right] \\
		 -\sum_{i,j} \gamma_{ij} \left( \left[\hat{\sigma}_i^{+},\hat{\sigma}_j^{-}\rho_{\lbrace a \rbrace } \right]+
		 \left[\rho_{\lbrace a \rbrace }\hat{\sigma}_i^{+},\hat{\sigma}_j^{-} \right] \right)
		  - \frac{1}{\hbar^2} \sum_{i,j} \Big( \left[ \hat{\sigma}_i^{+}, \left[ \hat{\sigma}_j^{-},\rho_{\lbrace a \rbrace }\right] \right] \int_0^t dt' {\bf d}_i \overleftrightarrow {\bf W}({\bf r}_i,{\bf r}_j,t,t' ){\bf d}_j e^{ i\omega_j (t-t')}\\
		 + \left[ \hat{\sigma}_i^{-}, \left[ \hat{\sigma}_j^{+},\rho_{\lbrace a \rbrace }\right] \right] \int_0^t dt' {\bf d}_i \overleftrightarrow {\bf W}({\bf r}_i,{\bf r}_j,t,t' ){\bf d}_j e^{-i\omega_j (t- t')} \Big).
\end{multline}
\end{widetext}
Here $\Omega_i({\bf r})=\Omega_i e^{-i{\bf k}_L {\bf \cdot r}}$,  $ \Omega_i={\bf d}_i\cdot \bm{\mathcal{E}}/2\hbar$ is the Rabi frequency.

The last sum in this equation is responsible for the various exchange channels between the populations and coherences induced by multiple scattering of photons. This follows from the operator part, i.e., the commutations of the projectors with density matrix. The time integral originates from the term commented on under equation (\ref{W1}). In this paper we will assume that all photoluminescence leaves the emitter system immediately with no chances to return. In other words, we state the conditions when we can ignore the multiple scattering. The choice of conditions that would allow approximate evaluation of time integrals deserves special attention. Solving the master equation with the feedback loop is a subject of our separate study and will be reported in another paper.

Thus, we have prepared a new master equation that describes the evolution of two near-identical two-level emitters driven by an incident field. This master equation is purely universal in terms of the incident plane wave geometry and positions of the emitters. For large interatomic separations, i.e., for $kr_{12}\gg 1$ and $\gamma_{12}=\Omega_{12}=0$ in accordance with their dependence on $r_{12}$,  the cooperative coupling is very weak or can be assumed zero and  the master equation is reduced to the Lindblad master equations for single atoms.

Note that we obtained the master equation for cooperative system after we started with the conventional electric dipole Hamiltonian and performed the other conventional steps, i.e., Born-Markov and the rotating wave approximations.  No by-hand terms were needed as dipole-dipole coupling and collective relaxation are organized by the BBGKY from the electric dipole potential for the interaction between the emitters and the quantized field.

\section{Emission intensity and excitation spectra}\label{Sec5}
In this section we are to calculate the total photoluminescence intensity from the coupled emitters as function of the pump frequency.  We will use this dependence to plot photoluminescence excitation spectra for different arrangements of the excitation and detection facilities.  From the literature we know that the emission intensity has been conventionally calculated upon the following reasoning: it is determined from the one-time first-order correlation function of the electromagnetic field at point ${\bf r}$  in the far-field zone of the radiation emitted by the atomic system \citep{AgarwalBook}. However we have at hand a new tool which may help us make a shortcut to the equation for the total photoluminescence intensity. It was shown in the previous section that BBGKY gives birth to field correlation tensor $\overleftrightarrow{\bf W}({\bf r}, {\bf r}', t, t')$.  Being a natural internal part of our system it may have different physical interpretations along with different values of its arguments.   If taken at one point $ ({\bf r},t) $, where $ {\bf r} $ is the position of the "observer" and $t$ is the time of registration,  the tensor would determine the amount of photoluminescence signal captured by the detector.  Thus, the electromagnetic field correlation tensor defined in (\ref{W1}) taken at the point $ ({\bf r}, t) $ allows us to calculate the total radiation intensity. One may combine the definition in equation (\ref{W1}) and the Poynting vector to show that ${\bf I}({\bf r},t)\sim (c/4\pi)\overleftrightarrow{\bf W}({\bf r}, {\bf r}, t, t)$.  Equation 18 can be evaluated if we use the formal solutions for  $ \rho_f $ (\ref{roF}) and $ g_ {ff '} $ (\ref{gFF}) from the equations 8 and 10 respectfully.  By completing two more steps,  which are the summation over the photonic modes and applying the Markov approximation, i.e.  $\rho_{\lbrace a \rbrace }(t')=\rho_{\lbrace a \rbrace }(t)$, we come to the expression for the two-point tensor:
\begin{multline*}
\overleftrightarrow {\bf W}({\bf r},{\bf r}',t,t') \\
=\int_0^{t'} d\tau Tr_{\lbrace a \rbrace } \Bigg\lbrace \sum_i \overleftrightarrow{\bf G}({\bf r}-{\bf r}_i,t-\tau)\widehat{\bf d}_i(\tau) \otimes \widehat{\bf \Phi} ({\bf r}',t') \\
+ \widehat{\bf \Phi} ({\bf r},t) \otimes \sum_i \widehat{\bf d}_i(\tau) \overleftrightarrow{\bf G}({\bf r}'-{\bf r}_i,t'-\tau) \Bigg\rbrace .
\end{multline*}
Here one can see that the solution for $\overleftrightarrow {\bf W}({\bf r},{\bf r}',t,t') $  produces $\widehat{\bf \Phi} ({\bf r},t)$, which contains $\overleftrightarrow {\bf W}({\bf r},{\bf r}',t,t')$ itself according to (\ref{Phi}).  It means that certain contributions will be reoccurring on a loop principle producing higher order terms.  On the other hand, our current objective is to extract and study pure photoluminesce process.  By looking again at (\ref{Phi}) and (\ref{roA}) one can clearly see that there is only a single term in $\widehat{\bf \Phi} ({\bf r},t)$ that produces the radiative relaxation operator in the master equation (\ref{roA}).  Consequently, it is sufficient to keep only this part of the complete $\widehat{\bf \Phi} ({\bf r},t)$ when substituting into the tensor $\overleftrightarrow {\bf W}({\bf r},{\bf r}',t,t')$ to describe spontaneous emission:
\begin{multline*}
\widehat{\bf \Phi}_\mathcal{L} ({\bf r},t) = \sum_j\int_0^{t} dt' \bigg \lbrace \overleftrightarrow{\bf G}^{+} ({\bf r}-{\bf r}_j,t-t')\widehat{\bf d}_j (t') \rho_{\lbrace a \rbrace } \\
+ \overleftrightarrow{\bf G}^{-} ({\bf r}-{\bf r}_j,t-t') \rho_{\lbrace a \rbrace }\widehat{\bf d}_j (t') \bigg \rbrace,
\end{multline*}
Substitution of the solution for the function $\hat{\bf \Phi}_\mathcal{L}  ({\bf r},t)$ and taking $r'\to r, t'\to t$ must be followed with applying the rotating wave approximation. This brings us to a new tensor that we refer to as intensity tensor $\overleftrightarrow {\bf I}({\bf r},t)$:
\begin{equation*}
\overleftrightarrow {\bf I}({\bf r},t) =\frac{c}{4\pi}\overleftrightarrow {\bf W}({\bf r},{\bf r},t,t)=\sum_{i,j} \langle \hat{\sigma}_i^{+} \hat{\sigma}_j^{-} \rangle \overleftrightarrow{\bf I}_{ij} ,
\end{equation*}
which is represented in the form of components $\overleftrightarrow{\bf I}_{ij}$ with factors that are the corresponding elements of the atomic density matrix $\langle \hat{\sigma}_i^{+} \hat{\sigma}_j^{-} \rangle \equiv Tr_{\lbrace a \rbrace } \lbrace\hat{\sigma}_i^{+} \hat{\sigma}_j^{-} \rho_{\lbrace a \rbrace }\rbrace$. The tensor components can now be written in the form where the corresponding time integrations have been performed using the same steps as in previous section.  It produces the Fourier transforms of the Green's tensors:
\begin{multline}\label{inten}
 \overleftrightarrow{\bf I}_{ij}= \frac{c}{\pi}{\bf d}_i^{*}\overleftrightarrow{\bf G}^{*} ({\bf r}-{\bf r}_i,\omega_i)\otimes\overleftrightarrow{\bf G} ({\bf r}-{\bf r}_j,\omega_j){\bf d}_j.
\end{multline}
Here, the spatial arguments describe the distances from each emitter to the observer. It is beneficial for further calculations to modify the Green's tensor as follows:
\begin{equation*}
\overleftrightarrow{\bf G} ({\bf r-r}_i, \omega_i) = \overleftrightarrow{\bf G} ({\bf r}, \omega_i) e^{-ik_i r_{ij}({\bf n}\cdot {\bf n}_{ij})/2} ,
\end{equation*}
where $ r_{ij}=\vert {\bf r}_i-{\bf r}_j\vert=r_{ji}$, ${\bf n}_{ij}={\bf r}_{ij}/r_{ij} =-{\bf n}_{ji}$ (note that ${\bf n}_{ii}=0$),  $ k_i=\omega_i/c$. Equation ({\ref{inten}) is general as it allows a solution for arbitrary orientations of the emitting dipoles.  However, for the problem stated in section \ref{ME} we may assume that the dipole moments are collinear and represent the vector of the dipole moment in the form ${\bf d}_i={ d}_i{\bf n}_d$.
Thus, we get:
\begin{equation} \label{int}
 \overleftrightarrow{\bf I}({\bf r},t) = \sum_{i,j} \langle \hat{\sigma}_i^{+} \hat{\sigma}_j^{-} \rangle \overleftrightarrow{\bf I}_{ij} e^{i\phi_{ij}+i(k_i-k_j)r },
\end{equation}
where $\phi_{ij}=r_{ij} k_0  ({\bf n}\cdot {\bf n}_{ij})=r_{ij} k_0 \cos{\varphi}$ is the observation parameter associated with the angle between vectors ${\bf n}_{ij}$ and $ {\bf n} = {\bf r} / r $ and
\begin{eqnarray*}
\overleftrightarrow{\bf I}_{ij} =&& \frac{\omega_i^2\omega_j^2 {d}_i{d}_{j}}{\pi r^2 c^3}\Big( ({\bf n}_d \otimes {\bf n}_d) \\
&&-({\bf n}_d \otimes {\bf n}+{\bf n} \otimes {\bf n}_d)({\bf n}_d \cdot {\bf n})+({\bf n} \otimes {\bf n})({\bf n}_d \cdot {\bf n})^2 \Big).
\end{eqnarray*}
In order to get a scalar expression for the intensity in the desired direction with respect of the orientation of the emitting dipole moments it is necessary to evaluate the values ${I}_{ij} = {\bf n}_{x_1} \overleftrightarrow{\bf I}_{ij} {\bf n}_{x_1} + {\bf n}_{y_1} \overleftrightarrow{\bf I}_{ij} {\bf n}_{y_1}$. Here vectors $n_{x_1}$ and $n_{y_1}$ are perpendicular to the vector directed at the observer and orthogonal to each other. Thus, we get a scalar value:
\begin{eqnarray} \nonumber
{I}_{ij} = \frac{\omega_i^2\omega_j^2 { d}_i{ d}_{j}}{\pi r^2 c^3} \sin^2{\beta},
\end{eqnarray}
where we introduce $ \beta $ as the angle between observation direction $ {\bf r} $ and the direction of the dipole moments $ {\bf d}_i $.

For a final evaluation of the total intensity we need the solutions for density matrix elements. The equations for these follow from the master equation  (\ref{roAA}). The master equation gives 15 kinetic equations describing the evolution of the atomic variables (components of density matrix).  Our goal is to find steady-state intensity so we put $\dot{\rho}_{\lbrace a\rbrace}=0$.  It is favourable to rewrite the elements of the atomic density matrix using the collective states representation \cite{Dicke1954}:
\begin{eqnarray*}
&\vert G \rangle=\vert g_1g_2 \rangle\\
&\vert S \rangle=1/\sqrt{2}(\vert e_1g_2 \rangle+\vert g_1e_2 \rangle)\\
&\vert A \rangle=1/\sqrt{2}(\vert e_1g_2 \rangle-\vert g_1e_2 \rangle)\\
&\vert U \rangle=\vert e_1e_2 \rangle
\end{eqnarray*}
In the collective states basis, the two-atom system behaves as a single four-level system with the ground state $\vert G \rangle$, the upper state $\vert U\rangle$, and two intermediate states: the symmetric state $\vert S\rangle$ and the antisymmetric state $\vert A\rangle$. The energies of the intermediate states depend on the dipole-dipole interaction.

In this way,  the total emission intensity from a two-particle cooperative system is written as
\begin{multline} \label{Intfin}
{\bf I}({\bf r})=\frac{I_{11}+I_{22}}{2}(2\rho_{UU}+\rho_{SS}+\rho_{AA})-2I_{12} \sin{\alpha}\mathrm{Im} \rho_{SA}\\+I_{12}\cos{\alpha}(\rho_{SS}-\rho_{AA})+(I_{11}-I_{22})\mathrm{Re} \rho_{SA},
\end{multline}
where $\alpha=ir_{ij} k_0  \cos{\varphi}+i(k_1-k_2)r$.

\begin{figure*}
  \includegraphics[width=0.3\textwidth]{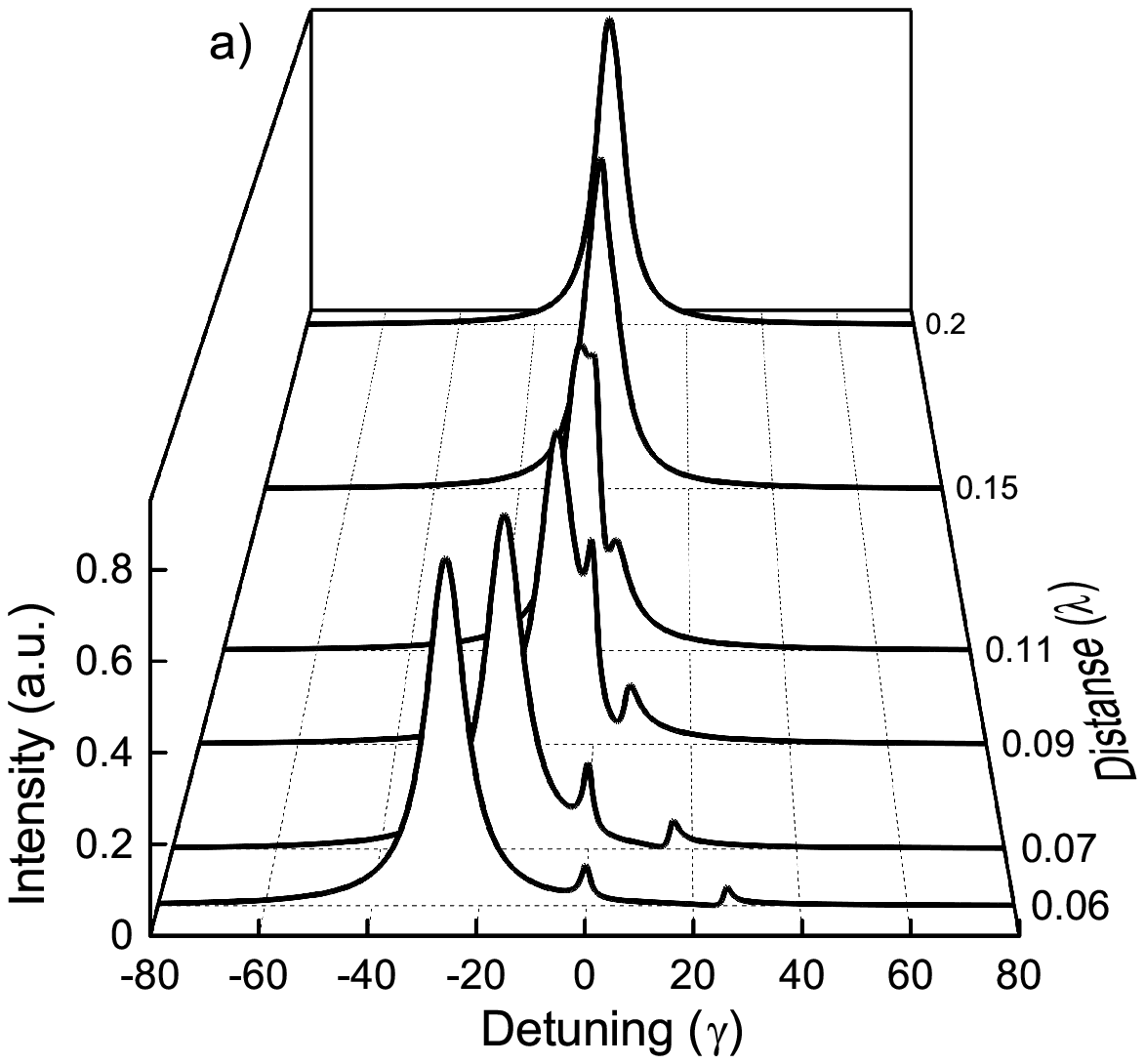}
   \includegraphics[width=0.3\textwidth]{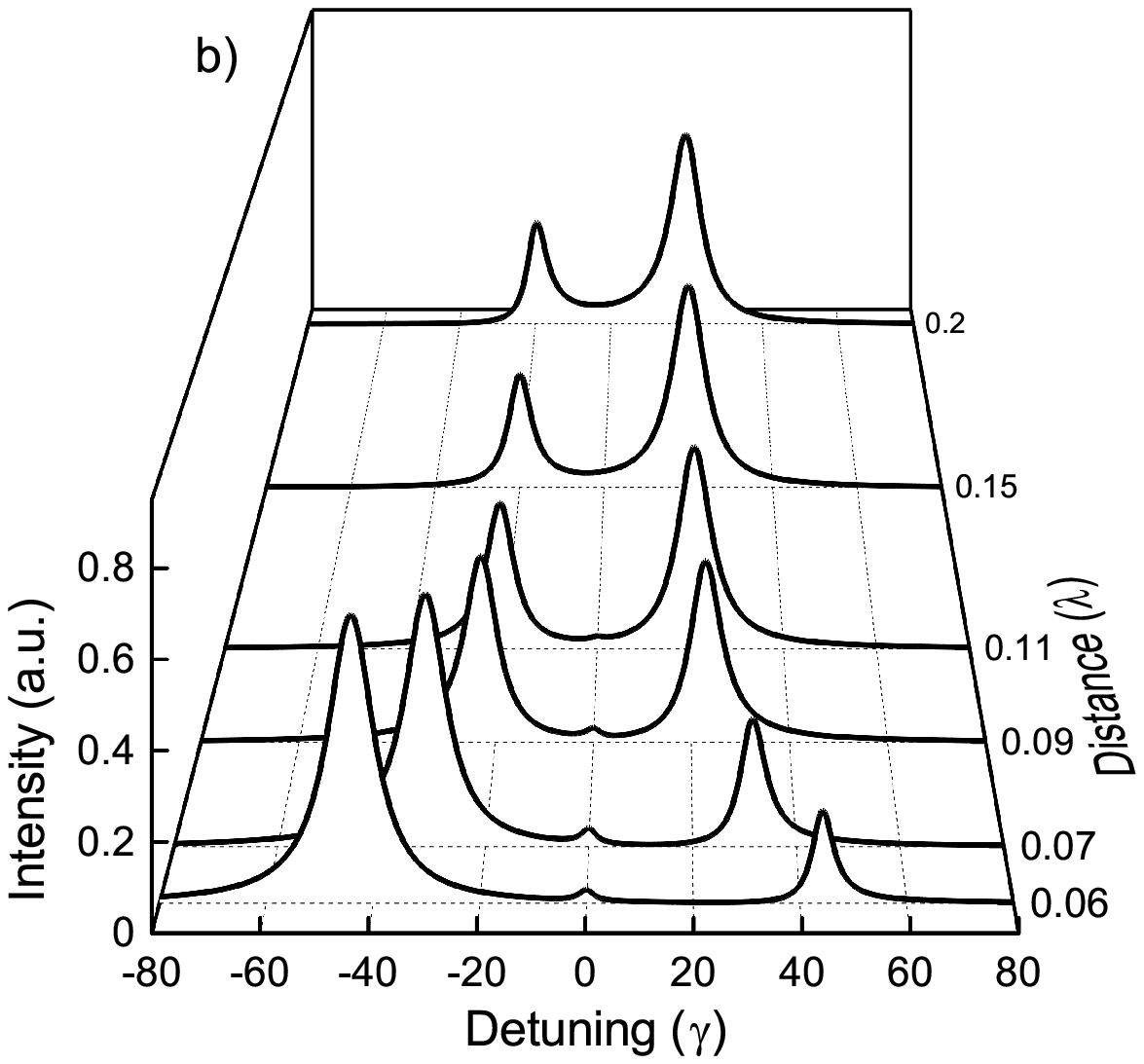}
    \includegraphics[width=0.3\textwidth]{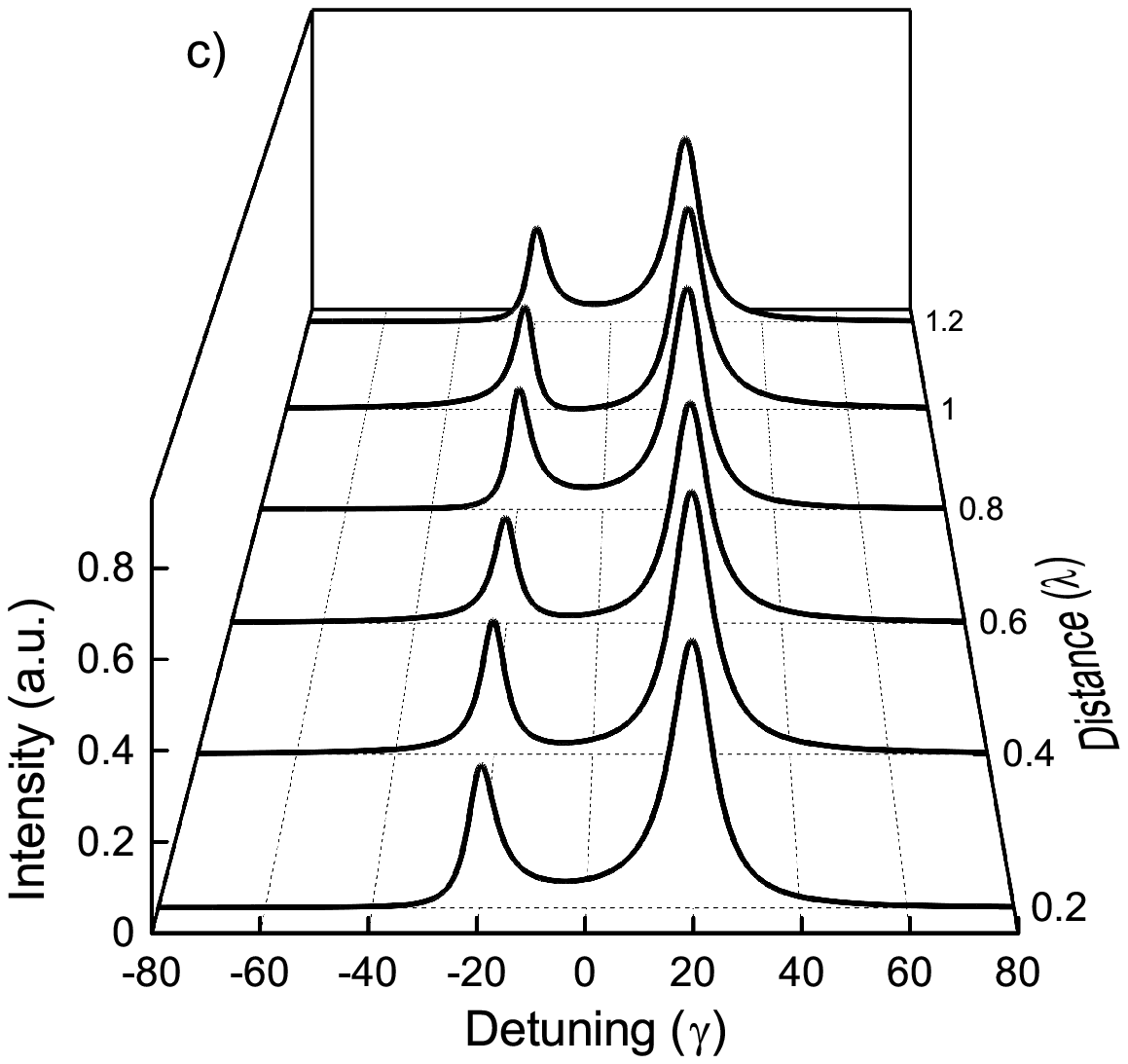}
  \caption{PLE spectra of two identical atoms (a) with $\Delta_{\omega}=0$ $d_1=d_2$,  and non-identical atoms (b) and (c) with $\Delta_\omega = 20$,  $d_2 = 1.5d_1$, for different distances between them.  Other parameters: $\Omega_1=2$, $\Omega_2=\Omega_1 d_2/d_1 $, $\xi=\pi$, $\theta=\pi/2$, $ \varphi=0$. }\label{Fig2}
\end{figure*}

Our objective is to investigate the photoluminescence excitation (PLE) spectra, i.e., the steady-state emission intensity as a function of the detuning $\Delta=(\omega_0-\omega_L)$,  where $\omega_0=(\omega_1+\omega_2)/2$. Registration of PLE spectra is a common experimental technique implemented with tunable lasers.
First, we will compare the entanglements occurring between two identical emitters with the entanglements possible for two near-identical emitters, i.e., with individual transition frequencies and moments. Since entanglement is enforced by the dipole-dipole coupling and collective decay it is of special interest to see how the PLE spectra are changing with the distance between the emitters. The incident beam parameters are also important in this regard. We will use the concept of observing photoluminescence from emitters with fixed positions when excitation may be subject to choice. Figures \ref{Fig2}(a)-(c) show the PLE spectra depending on the separation between the particles (in terms of resonance wavelength of one of the emitters $\lambda_1=\lambda$). The rear spectral profiles in Figs.\ref{Fig2}(a)-(b)  demonstrate weak entanglement when the separation is large enough to let the emitters act independently ($\Omega_{12}$ becomes small). For identical atoms in Fig.\ref{Fig2}(a) one can see a single aggregated peak corresponding to the resonance frequency of both emitters. For non-identical atoms in Fig.\ref{Fig2}(b) one finds two distinct peaks at the frequencies corresponding to each resonance.  Figure \ref{Fig2}(c) illustrates the situation when the emitters are at the distances when dipole-dipole coupling and collective decay are already negligible  small. In this case one can clearly see the two peaks which are largely individual PLEs. However when the excitation is tuned between the transitions we see the fluctuations of the resulting interference between individual emissions. This occurs due to non-vanishing matrix elements in (\ref{Intfin}). The aparency of this interference is determined by  angles $\xi$ and $\varphi$.

For smaller distances the spectra profiles become notably different (the front stage profiles in Figs.\ref{Fig2}(a)-(b)). Convergence of the atoms split the picture into three peaks. The three peaks demonstrate the well-known configuration of the cooperative resonances, observed experimentally in \cite{Hettich2002}. As was shown, the central peak corresponds to the simultaneous absorption of two photons and subsequent radiation of the pair of correlated photons. It is evidenced by photon-bunching in the autocorrelation function, which was also measured in the experiment.  The side peaks correspond to the collective states $\vert S\rangle$ and $\vert A \rangle$, where $\vert S\rangle$ is superrariative and $\vert A \rangle$ is subradiative. The only so-far successfully observation of the cooperative triplet  \cite{Hettich2002} indeed has brought the evidence of the physical picture we simulate in out study. They measured the PLE spectra from two terrylene molecules embedded in a para-terphenyl crystal. Terrylene molecules are very well studied and known to demonstrate wide inhomogeneous broadening. The emitters in \cite{Hettich2002} were claimed to have different transition frequencies and be separated by 12 nanometers. Besides, the conditions of the experiment were such that the authors suggested the two-level atom model was a good approximation for terrylene with regard to dipole-dipole coupling.  Figure \ref{Fig9} shows how the spectra changes under different excitation intensities.  It is seen that the central peak corresponding to two-photon process arises at certain incident power and increases with the increasing intensity. This behaviour is in a good qualitative agreement with the PLE plots in  \cite{Hettich2002}.  The spectral positions of the cooperative resonances, the ratios between the peak maximums and the widths are generally the same in both the experiment and the theory. However we have prepared the theory to show in detail how the PLE shape is dependent on the geometry of the problem. Moreover,  for a quantitative comparison with the measurements it may be also necessary to take care of some special issues like the following: 1) the choice of the appropriate approximation for the emitters; 2) the role of the conic collection of photoluminescence photons; and 3) possible effects of the data processing.
\begin{figure}
  \includegraphics[width=0.4\textwidth]{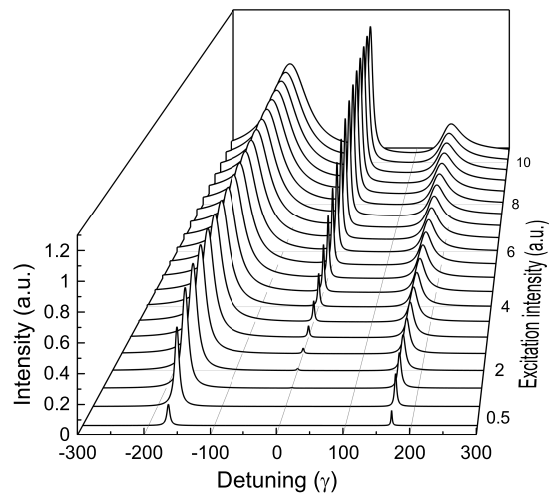}
  \caption{PLE spectra for different excitation strengths with $\Delta_\omega=100$, $d_1=1.5d_2$, $\xi=\pi/2$, $\varphi=\pi/2$, $r_{12}=0.04\lambda$. }\label{Fig9}
\end{figure}
\begin{figure}
 \includegraphics[width=0.35 \textwidth]{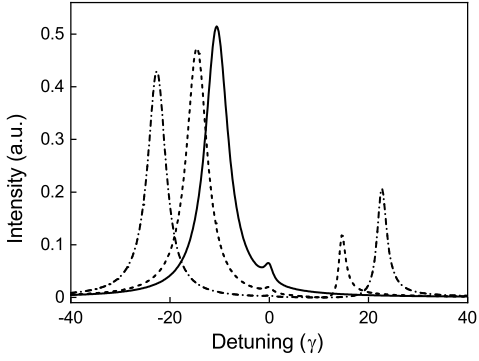}
 \caption{PLE spectra for identical atoms (solid line), and non-identical atoms with $\Delta_\omega=10$, $d_1=d_2$ (dashed line) and $\Delta_\omega=20$, $d_1=d_2$ (dash-dotted line). Other parameters: $\Omega_1=\Omega_2=2$, $\xi=\pi/2$,  $\theta=\pi/2$, $ \varphi=\pi/2$,  $r_{12}=0.08\lambda$.}\label{Fig3}
 \end{figure}

Now let us look at how geometry of the spatial arrangement of the emitters against the excitation and detection modify the PLE profiles.  First, we consider the case when the excitation is perpendicular to the interatomic axis, i.e.,  ${\bf k}_L \perp {\bf r}_{12}$.  It is interesting in terms of comparison with the results known in the literature for two identical emitters. That variant ($\omega_{1}=\omega_{2}$, ${\bf d}_{1}={\bf d}_{2}$) was shown to demonstrate photoluminesce intensity with only two distinct peaks at $\Delta=0$ and $\Delta=-\Omega_{12}$ (solid line in Fig.\ref{Fig3}).  It means that at this geometry the peak corresponding to the antisymmetric state $\vert A \rangle$ is not visible.  At the same time, for different atoms the spectra exhibits three peaks at $\Delta=0$ and $\Delta\simeq\pm\sqrt{\Omega_{12}^2+\Delta_\omega^2}$, $\Delta_\omega=(\omega_1-\omega_2)/2$ (dashed and dash-dotted lines in Fig.\ref{Fig3}).

The distinctions between the emitters may as well be vividly demonstrated at another variant of geometry.  The key feature is to observe photolimescence from different angles (Fig.\ref{Fig4}).  For ${\bf r}\perp {\bf r}_{12}$ ($\varphi=\pi/2$) the fluorescence intensity for identical emitters exhibits two peaks centered at $\Delta=0$ and $\Delta=-\Omega_{12}$ and a dip at $\Delta=\Omega_{12}$ (solid line in Fig. \ref{Fig4}). For $\varphi<\pi/2$ this dip changes to a peak (dashed line in Fig. \ref{Fig4}). The situation is different when atoms are non-identical. The fluorescence intensity exhibits three peaks for all observation parameters with small changes in peak height(inset in Fig. \ref{Fig4}).

\begin{figure}
  \includegraphics[width=0.35\textwidth]{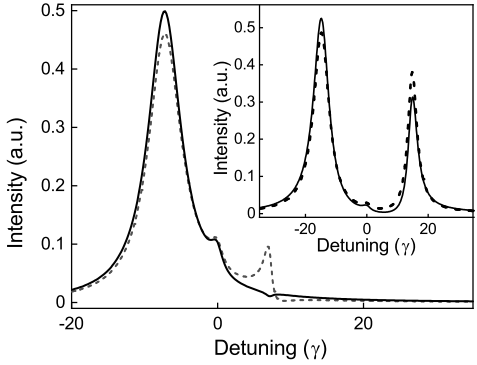}
 \caption{PLE spectra for identical atoms for different directions of observation:  $\varphi=\pi/2$ (solid line),   $\varphi=0$ (dashed line). The inset show the same picture for non-identical atoms with $\Delta_\omega=10$, $d_1=d_2$. Other parameters: $\Omega_1=\Omega_2=2$, $\xi=\pi$,  $\theta=\pi/2$,  $r_{12}=0.09\lambda$.}\label{Fig4}
\end{figure}
\begin{figure}
  \includegraphics[width=0.3\textwidth]{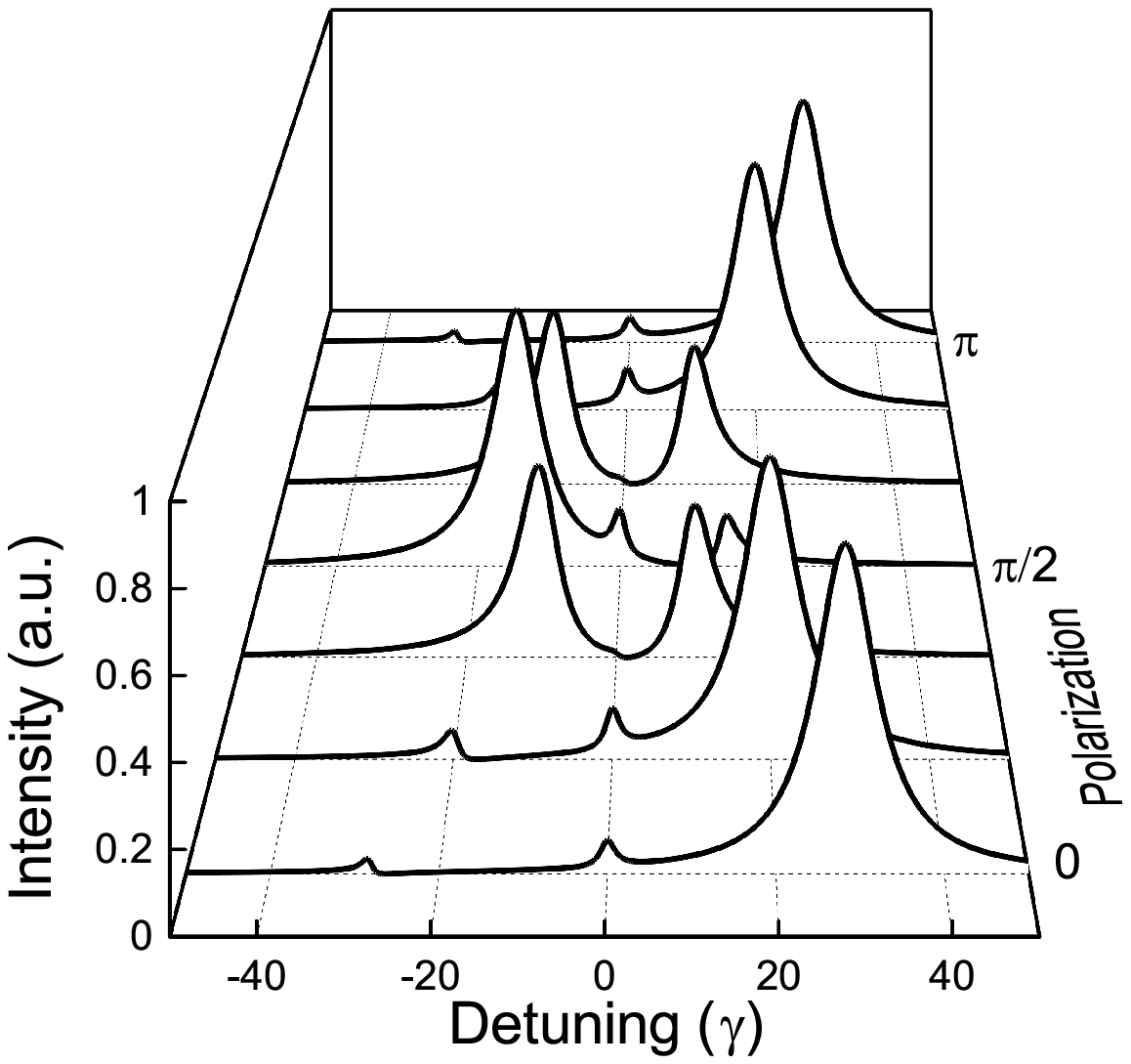}
  \caption{Migration of the PLE spectra upon rotation of the polarization of an external field for $\Delta_\omega=10$, $d_1=d_2$, $\xi=\pi/2$, $\varphi=\pi/2$, $r_{12}=0.08\lambda$, $\Omega_1=\Omega_2=2$. }\label{Fig5}
\end{figure}
\begin{figure}
  \includegraphics[width=0.4\textwidth]{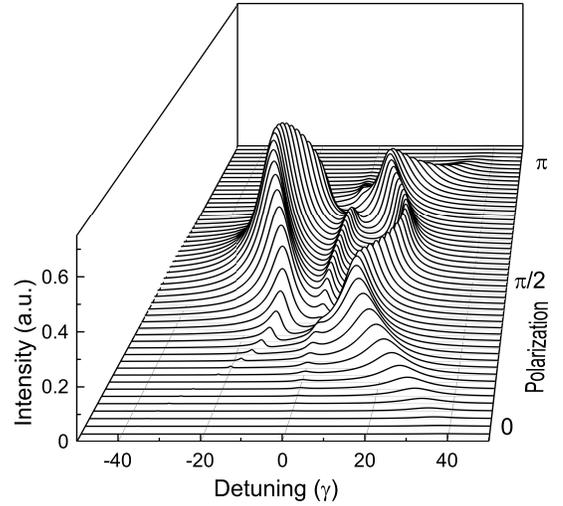}
  \caption{Migration of the PLE spectra upon rotation of the polarization of an external field for $\Delta_\omega=10$, $d_1=d_2$, $\xi=\pi/2$, $\varphi=0$, $r_{12}=0.08\lambda$, $\Omega_1=\Omega_2=2$. }\label{Fig6}
\end{figure}

From the analysis above it appears that two different emitters may happen to be entangled in the way that provides better conditions for efficient realisation of the subradiative state.  Experimentally one may hope to observe the asymmetric state emission because it has a greater spectral separation from the other cooperative resonances and notably stronger intensity.
\begin{figure}
\includegraphics[width=0.4\textwidth]{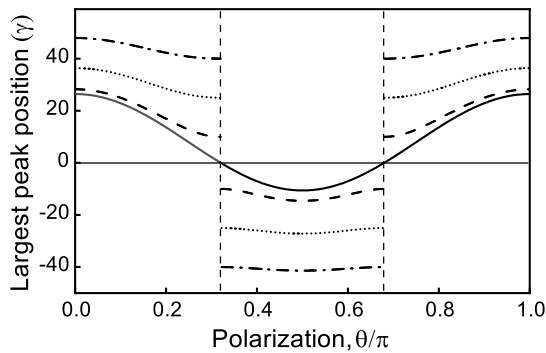}
\caption{Position of the maximum fluorescence peak relative to the central peak  as a function of polarization ($\theta$) for $\Delta_\omega= 0$ (solid line), $\Delta_\omega=10$ (dashed line), $\Delta_\omega = 25$ (dotted line), $\Delta_\omega=40$ (dashed-dotted line) for $r_{12}=0.08\lambda$. }\label{Fig7}
\end{figure}

Now we focus on the particular parameters of the incident field. It is clear that dipole-dipole coupling parameter is a strong function of angle $\theta$ as it is shown in Fig.\ref{Fig8} to change from zero to considerable absolute values.  Therefore is can be straightforwardly anticipated that the form of the PLE spectra strongly depends on the incident field polarisation, i.e., on the angle $\theta$ (see Fig.\ref{Fig1} and the corresponding comments in section \ref{2b}).  The actual relation between the configuration of the beam polariser and $\theta$ is to be calculated for a particular experimental setup.  For an illustration of the effect it is sufficient to choose the incidence such that $\theta$ is the actual rotation angle of ${\bf E}_L$ around ${\bf k}_L$.  The transformation of PLE pattern with polarization rotation were calculated for a fixed distance between emitters as shown in Fig.\ref{Fig5}.  This Figure shows how the spectral distance between the peaks appears to be shrinking with growing $\theta$ until the pattern becomes a two-peak PLE.  That is because at certain angle $\theta$ we have $\Omega_{12}=0$ and the distance between peaks is determined by the difference in the transition frequencies of the emitters. With further increase of $\theta$ the PLE is again a cooperative triplet in which the symmetric and antisymmetric peaks have exchanged their positions against the origin. From the quarter-turn of $\theta$ the PLE pattern starts to unwind back to the initial picture. Obviously, with a half-turn the positions of the peaks restore completely.  Figure \ref{Fig6} shows similar behaviour as in Fig.\ref{Fig5} but with finer discreteness., while the observation angle was chosen to be such that no emission is observable at both margins, i.e., $\sin \beta=0$ at $\theta=0,\pi$.  In this case one would see how a half-turn of the excitation polarization makes a photoluminescence signal grow from null, reach maxima and disappear.  For each excitation frequency the emission intensity will change in some particular manner. In other words, it may be an individual function of $\theta$.  The general picture appears as a gorge formed by the symmetric and asymmetric summits. The two narrow passes characterize the differences of the transition frequencies $\Delta_\omega$, while the width of the gorge at the quarter-turn, $\theta=\pi/2$, indicates the distance between the emitters. Note that the small central chine is peculiar for weak or moderate excitation powers. Another interpretation of the changed brought by turning the polarization of the excitation beam is the "spectral jump" of the symmetric (or asymmetric) maximum.  Figure \ref{Fig7} demonstrates the spectral migration of the corresponding maximum.  For identical emitters this migration produces the indissoluble trail. On the contrary, the emitters with different transition frequencies produce broken trails shown in Fig.\ref{Fig7}, where the gaps are regulated by $\Delta_\omega$.
These properties suggest that measuring the amplitude of spectral migration and the length of spectral jumps can help detection and characterization of paired emitters.

\section{Conclusion}
In conclusion, we have studied the steady-state photoluminescence from two near-identical emitters with dipole-dipole coupling. We considered two two-level particles with individual transition frequencies and transition moments. The excitation of the system was assumed as an incident continuous monochromatic beam of light with linear polarization. The geometrical arrangements of the excitation beam, positions of the emitters and the detector were included into the solutions. The calculations were performed using the formalism based on the BBGKY hierarchy of equations, adapted to solving light-matter interactions and spectroscopic problems. The von Neumann equation for the density matrix of a many-particle system was written with the conventional electric-dipole Hamiltonian for two emitters and photons. The excitation signal was set via the initial conditions for the photonic modes. The full von Neumann equation was expanded into the BBGKY chain which was later truncated into a closed system of equations for reduced density matrices  and correlation operators. The system of equations described the two-particle emitter, the photonic modes, and the emitter-field and intrafield correlations. It was treated with the use of the conventional approximations to give the cooperative master equation and the equation for the total photoluminescence intensity. The cooperative entanglement and radiative decay, as well as the dipole-dipole interaction appeared naturally in the master equation as the result of the decoupling procedures applied to the BBGKY.

The master equation and the intensity equation were used to perform computer simulations of PLE spectra, i.e., the total emission against the excitation frequency. We showed the properties of the spectral shapes which can reveal the differences between the emitters and demonstrate the importance of the excitation-detection geometry relative to the emitting pair. The spectra were proved to fully agree with the previous thoeries and the experiment. Two non-identical emitters exhibit a pronounced cooperative PLE triplet in which the central peak corresponds to the excitation equally detuned from both emitters transitions while two side peaks correspond accordingly to the super- and sub-radiative states of the pair or the cooperative resonances. At the same time we have found new properties of photoluminescence from the near-identical emitters which are the following:

1. Non-identical emitters versus identical emitters need smaller separation to start showing a pronounced cooperative PLE triplet. However, the non-identical emitters are more favourable for experimental investigations because the spectral distance between the peaks in PLE is significantly greater while the antisymmetric peak (sub-radiative state) is substantially inflated. Besides, there are several configurations of the excitation-detection geometry for which the symmetric and antisymmetric peaks become comparable in both the intensity and the width.

2. At fixed incidence of the excitation beam and fixed  registration direction the rotation of the excitation beam polarisation may become a handy tool in either nanoscopy or recovering of the diversity of the emitters. We showed that there is spectral migration of the brightest peak in PLE that follows the changing polarisation. This can be a good indication of the presence and the strength of the dipole-dipole interaction as the migration occurs within a certain range.  Moreover, we found the existence  of the spectral pseudo-jump of the brightest peak which may serve as the measure of the difference in transition frequencies. In this case a PLE trail would have a point corresponding to a certain polarisation at which the dipole-dipole interaction vanishes and the spectral distance between symmetric and antisymmetric resonances is determined solely by the difference of the  transition frequencies.

Note that our results were obtained with the least possible order of the BBGKY decoupling approximation. The use of the higher order terms provided in the BBGKY would surely reveal fine details of the cooperative photoluminescence.  In this paper we have paved the direction to further important improvements of the theory.  A strong contribution may be expected from the field correlation tensor which is present in the full master equation. This contribution is generally non-vanishing because it appears in the same order of the BBGKY decoupling. It provides additional terms in kinetics of populations and coherences due to multiple scattering of photons.  It may be of special importance for particular conditions which provide confinement or trapping of the emitting pair. Our findings may be important for heading towards all-optical nanoscopy and construction of entangled systems to be used in quantum technologies.

\begin{acknowledgments}
This work was carried out among the State Contract of the Moscow Pedagogical State University (MPGU) “Physics of the perspective materials and nanostructures: basic researches and applications in material sciences, nanotechnologies and photonics” supported by the Ministry of Education of the Russian Federation (AAAA-A20-120061890084-9) in collaboration with the Centre of collective usage “Structural diagnostics of materials” of the Federal Research Center RAS “Crystallography and photonics”.
\end{acknowledgments}

\end{document}